\documentclass[10pt,conference]{IEEEtran}

\usepackage[utf8]{inputenc}
\usepackage{cite}
\usepackage[ruled,vlined]{algorithm2e}
\usepackage[T1]{fontenc}
\usepackage{microtype}
\usepackage{graphicx}
\usepackage{paralist}
\usepackage{tabularx}
\usepackage{balance}
\usepackage{multicol}
\usepackage{multirow}
\usepackage{pbox}
\usepackage{amssymb}
\usepackage{colortbl}
\usepackage{pifont}
\usepackage{xspace}
\usepackage[dvipsnames]{xcolor}
\usepackage{url}
\usepackage[TABBOTCAP]{subfigure}
\usepackage{rotating}
\usepackage{tikz}

\usepackage{booktabs}

\newcommand{\ie}{\textit{i.e.,}\xspace}
\newcommand{\aka}{\textit{a.k.a.,}\xspace}
\newcommand{\eg}{\textit{e.g.,}\xspace}
\newcommand{\etc}{\textit{etc.}\xspace}
\newcommand{\etal}{\textit{et al.}\xspace}

\newcommand{\secref}[1]{Section~\ref{#1}\xspace}

\newcommand{\figref}[1]{Fig.~\ref{#1}\xspace}

\newenvironment{myquote}%
  {\list{}{\leftmargin=0.1in\rightmargin=0.1in}\item[]}%
  {\endlist}

\include{macro}

\begin{document}

\title{An Empirical Study on\\Android-related Vulnerabilities}

\author{
\IEEEauthorblockN{
Mario Linares-V\'asquez$^1$, Gabriele Bavota$^2$, Camilo Escobar-Vel\'asquez$^1$
}
\IEEEauthorblockA{
$^1$ Systems and Computing Engineering Department, Universidad de los Andes, Bogot\'a, Colombia \\
$^2$ Faculty of Informatics, Universita della Svizzera Italiana, Lugano, Switzerland\\
m.linaresv@uniandes.edu.co, gabriele.bavota@usi.ch, ca.escobar2434@uniandes.edu.co
}
}

%\IEEEoverridecommandlockouts

%\IEEEpubid{\makebox[\columnwidth]{978-1-5090-1428-6/16/\$31.00~
%\copyright2016
%IEEE \hfill} \hspace{\columnsep}\makebox[\columnwidth]{ ICPC 2016, Austin, Texas}}

\maketitle

\begin{abstract}
Mobile devices are used more and more in everyday life. They are our cameras, wallets, and keys. Basically, they embed most of our private information in our pocket. For this and other reasons, mobile devices, and in particular the software that runs on them, are considered first-class citizens in the software-vulnerabilities landscape. Several studies investigated the software-vulnerabilities phenomenon in the context of mobile apps and, more in general, mobile devices. Most of these studies focused on vulnerabilities that could affect mobile apps, while just few investigated vulnerabilities affecting the underlying platform on which mobile apps run: the Operating System (OS). Also, these studies have been run on a very limited set of vulnerabilities. 

In this paper we present the largest study at date investigating Android-related vulnerabilities, with a specific focus on the ones affecting the Android OS. In particular, we (i) define a detailed taxonomy of the types of Android-related vulnerability; (ii) investigate the layers and subsystems from the Android OS affected by vulnerabilities; and (iii) study the survivability of vulnerabilities (\ie the number of days between the vulnerability introduction and its fixing). Our findings could help OS and apps developers in focusing their verification \& validation activities, and researchers in building vulnerability detection tools tailored for the mobile world.
\end{abstract}

\IEEEpeerreviewmaketitle

\section{Introduction}
\label{sec:intro}
In the last few years, mobile apps have powered a whole new economy that substantially impacted the software market. The cultural popularity of mobile devices, the new monetization/revenue models the apps' market propose, and the capillary distribution infrastructure represented by app stores, are only some of the driving factors making apps an attractive market for software developers. Also, the need for ``enterprise apps" that support startups or serve as a new front-end for traditional companies is pushing software-related professionals to embrace the mobile technologies \cite{VisionMobile:14}. 

From the users' perspective, mobile apps and devices are a mechanism for achieving ubiquity, allowing them to perform multiple tasks and daily activities from anywhere, and to always have available at the touch of their hands important/sensitive information. Consequently, the security of mobile apps and of the underlying platforms on which they run has become a big concern for researchers and practitioners, due to the impact that security issues affecting mobile platforms might have on the private life of individuals (\eg allowing to stole private files) as well as on companies (\eg allowing to intercept strategic business decisions) \cite{Ponemon:2014,Stefanko:2015,Anderson:2016}.

Recently, the impact of those vulnerabilities in everyday life has been more evident to the society due to public announcements of malware and vulnerabilities in mobile platforms that compromise sensitive information and/or computational resources in the affected devices.  In 2015 mobile malware reached a tremendous +153\% (Android) and +235\% (iOS) in the number of reported threats as compared to the previous year \cite{Anderson:2016}. Representative examples of mobile malware with notorious impact are games such as ``Cowboy adventure" and ``Jump chess" that infected about 1 million devices\cite{Beres:2015},  the Locker trojan \cite{Stefanko:2015,Anderson:2016} for Android, and the XcodeGhost malware that infected  40K+ apps from the Apple App Store \cite{Anderson:2016}. Also, according to the CVE details portal \cite{CVE-Android}, 125 and 523 vulnerabilities in the Android OS were reported in 2015, and 2016, respectively. One of those vulnerabilities is ``Stagefright" \cite{Burgess:2016,Nickinson:2015,Stagefright} that compromised 95\% of the Android devices in 2015.

As a contribution from the research community, substantial effort has been recently invested in the analysis and detection of malware and vulnerabilities at the applications level (see \eg \cite{Sadeghi:ICSE15,Bagheri:TSE15,Ahmad:MSR16}). However, (i) few works have focused on the vulnerabilities at the OS level \cite{Thomas:2015,Thomas:SPSM2015,Jimenez:QRS16, Bagheri:SJFAC,Cao:ACSAC15}, the underlying platform on which any app runs, and (ii) most of the studies have just focused on a limited number of vulnerabilities. 

In this paper, we present an empirical study aimed at analyzing from several different perspectives Android-related vulnerabilities, with a specific focus on those affecting the Android OS. In particular, we study (i) the types of vulnerability, (ii) the layers and subsystems from the Android OS affected by vulnerabilities, and (iii) the survivability of vulnerabilities (\ie the number of days between the vulnerability introduction and its fixing). While previous studies have focused the attention on a small set of vulnerabilities (\eg 11 in \cite{Thomas:SPSM2015}, 1 in \cite{Thomas:2015}, and 32 in \cite{Jimenez:QRS16}), we mined all the vulnerabilities (660) available in the official Android bulletins and the CVE-details portal up to November 2016.  The vast majority of our study has been carried out via manual analysis of vulnerability-related documents available on issue trackers, versioning system, official Android security bulletins, and information available on the National Vulnerability Database\cite{NVD-RSS}. 

Knowing the types of security vulnerabilities affecting the Android devices and their characteristics can help to guide (i) apps developers, in focusing their verification \& validation activities toward the identification of the most frequently reported types of vulnerability,  (ii) researchers, in investing in vulnerabilities detection tools targeting the most diffused types of Android-related vulnerabilities (thus being particularly valuable to increase the security of Android devices), and (iii) language/API developers, to design/improve mechanisms for secure coding of Android apps and the underlying platform. %This paper contributes to the community by reporting the most critical layers/subsystems in the Android OS (in terms of security vulnerabilities), the most frequent types of vulnerabilities, and by highlighting the prevalence of Android OS vulnerabiliries when using as reference the survavility time. %GABRIELE: I think this is redundant wrt the next paragraph

As a result of our study we defined a detailed taxonomy of vulnerabilities affecting Android devices (\figref{fig:types}) based on the Common Weaknesses Enumeration~\cite{CWE}, and identified the layers/subsystems of the Android OS mostly impacted by vulnerabilities (\figref{fig:heatmap}).  We found that the hardware drivers in the lowest level of the  Android OS stack (\ie, linux kernel), and the native libraries are the layers mostly impacted by security vulnerabilities, and the lack of secure coding practices for restricting operations in the bounds of memory buffers is the main source of vulnerabilities. In addition, we found that Android-related security vulnerabilities survive for very long time (at least 724 days, on average).

%\textbf{Structure of the paper}. We provide some background on the Android OS and discuss the related literature in \secref{sec:background}. \secref{sec:design} presents the design of our study, while our findings are discussed in \secref{sec:results}. \secref{sec:threats} lists the threats that could affect the validity of our results. Finally, \secref{sec:conclusion} outlines the conclusions.

\section{Background and Related Work}
\label{sec:background}
%\MARIO{TODO intro paragraph}

%\subsection{The Android OS Software Stack}
The Android OS is an open source mobile OS developed by Google and based on the Linux Kernel. It is composed of a set of architectural layers that follows a software stack model, having the Linux Kernel as the foundation, and an Applications layer as the closest interaction point for the end users. Each layer is composed of subsystems/components mostly implemented in Java and C/C++. %; for instance the ``Libraries" layer contains an Android's standard C library called Bionic, and the ``Runtime" layer includes an independent Java Library implementation (Apache harmony %or an OpenJDK-based implementation for newer OS versions).  
Some of those components are developed by third-party contributors of the Android open source project (AOSP), such as original equipment manufacturers (OEM) and Linux contributors.

%\figref{fig:android-stack} depicts the Android OS software stack (including predominant implementation languages) and we briefly describe the Android OS layers as in the following:

The Android OS stack is composed by the following layers:

{\tt Applications:} software running on the device that uses the Android APIs to implement specific features, like geo-localization. The components in this layer are the mobile ``apps" we use daily such as Browser, Calendar, and Settings; these apps are mostly written in Java. 

{\tt Android Framework:} provides apps (and developers) with the building blocks and common tasks required for exposing/using device- and Android-specific features such as managing UI elements and sensors. The Android Framework contains the Android APIs and Android managers (\aka services); examples of these services  are the View System and the Activity Manager. This layer is implemented in Java.

{\tt Runtime:} contains the Virtual Machine (Dalvik/ART) and the core libraries required for the execution of apps and services on the device. Runtime is required for ensuring apps portability across different devices. Examples of the core libraries in the Runtime layer are the independent implementation of Java used by Android and the Bouncy castle library.

{\tt Native Libraries:} provide low level functionality and computational intensive services required by the Android Framework and the Runtime, such as the Bionic libc library, the WebKit browser engine, OpenGL, SSL, and the Media Framework. The libraries are written in C/C++.

{\tt Hardware Abstraction Layer (HAL):} it is the bridge between the high level representations of the hardware used in the libraries, and low level representations used by the kernel. It is a set of interfaces for hardware-specific software that needs to be implemented by OEMs and hardware manufacturers. Components in the HAL are written in C/C++.

{\tt Linux Kernel:} it provides the Android OS with core OS systems infrastructure, a security model,  networking, and memory and process management, among the others. Android uses a modified version of Linux tailored to mobile devices.%; this tailored version includes changes/enhancements such as the Android Binder, Logger, ashmem, power management, wakelocks, and mechanisms for memory management (e.g., Android Low Memory Killer).

For more details of the Android OS architecture, we point the interested reader to the following sources: \cite{Brady:2008,Google:Platform,Drake:2014}.

%\begin{figure}[t]
%\begin{center}
%\includegraphics[width=0.7\linewidth]{img/android-stack}
%\caption{Architectural layers in the Android OS software stack}\vspace{-0.5cm}
%\label{fig:android-stack}
%\end{center}
%\end{figure}

\subsection{Malware and Vulnerabilities}
The wide and rapid adoption of Android-based devices in the last years has motivated the usage of Android apps to support a broad range of daily activities. In that sense, being the most popular mobile platform makes it an attractive target for security attacks  \cite{Huang:CCS15}. In fact, the number and complexity of the attacks to Android-based systems is increasing drastically \cite{Huang:CCS15}; since 2008, Android-related vulnerabilities have been reported including critical issues such as the ``Heartbleed"\cite{heartbleed} flaw in the SSL library, and the ``Stagefright" flaw in the media framework \cite{Stagefright,Nickinson:2015,Burgess:2016} that has infected 95\% of the Android devices in 2015.  As a consequence, the industry has improved the security mechanisms and services in the Android ecosystem \cite{Google:20016} and designed mobile-specific malware detectors. 

Researchers have also contributed to improve the security of the Android ecosystem by analyzing security vulnerabilities and proposing improvements to current security models \cite{Bagheri:SJFAC,Huang:CCS15,Xu:ACM16,Cao:ACSAC15,You:ICSE16,Wang:CCS16,Wu:CCS13,Ahmad:MSR16,Zhou:SSP12,Sufatrio:ACM2015,Sadeghi:TSE16,Fahl:CCCS12,Fahl:CCCS12,Zuo:CCS15,Backes:CCS16,Kantola:SPSM12}; however, while the focus of the academic research has been the security of the applications---the closest component to the user---,  the core of the Android ecosystem (\ie the Android OS) has received little attention. 

\subsubsection{Security in Android Applications} Android malware and vulnerabilities in Android apps are characterized by a novel set of flaws that exploit user level weaknesses and the issues in security mechanisms of the Android OS. For instance, Android-specific attacks include (i) privileges/permissions escalation through pairs of infected apps that exploit inter-application communication or misconfigured apps\cite{Kantola:SPSM12,Sbirlea:IBM13,Sadeghi:ICSE15,Bagheri:TSE15,Ahmad:MSR16}, (ii) applications tapjacking/hijacking by apps repackaging and substitution\cite{You:ICSE16},  (iii) information leaking through covert channels\cite{Gasior:ICCS12,Novak:UBICOMP15},  (iv) SSL vulnerabilities in hybrid \cite{Zuo:CCS15} and native apps\cite{Fahl:CCCS12}, (v) security issues introduced by third party libraries\cite{Backes:CCS16}, and  (vi) security issues introduced by OS customizations\cite{Wu:CCS13}. These novel attacks, in addition to classic security attacks induced by malware (\eg DoS), have been widely studied by the community and several approaches have been proposed for their detection and mitigation, such as TaintDroid\cite{Enck:USENIX10}, COVERT \cite{Sadeghi:ICSE15,Bagheri:TSE15}, FlowDroid\cite{Arzt:PLDI14}, MudFlow\cite{Avdiienko:ICSE15}, Chabada\cite{Gorla:ICSE14}, Q-Floid\cite{Castellanos:SGCRC16}, and AppInspector\cite{Gilbert:MCS11}. Other resources, like the Android Malware Genome Project (Malgenome) \cite{malgenome}, aim at characterizing Android malware families by describing installation methods, activation mechanisms, and malicious payloads; the Malgenome project includes 1,200 malware samples collected from August 2010 to October 2011. For more details we refer the interested reader to the works by Zhou \etal \cite{Zhou:SSP12} and Sufatrio \etal \cite{Sufatrio:ACM2015} that widely describe Android malware and detection techniques, and a recent work by Sadeghi \etal \cite{Sadeghi:TSE16} presenting a survey of static analysis techniques for detecting Android malware.

\subsubsection{Android OS Vulnerabilities} Previous studies focused on the analysis of specific components of the OS and their security issues. Bagheri \etal \cite{Bagheri:SJFAC} analyzed  the vulnerabilities of the permission system in the OS; Cao \etal \cite{Cao:ACSAC15} analyzed input validation mechanisms in the services/managers of the {\tt Android Framework}; Huang \etal \cite{Huang:CCS15} found 4 vulnerabilities (\aka Android Stroke Vulnerabilities) in two services of the {\tt Android Framework} (\ie Activity and Window Manager) that can be used for DoS attacks and for inducing OS soft-rebooting;  Wang \etal \cite{Wang:CCS16} also analyzed the {\tt Android Framework} layer and found six unknown vulnerabilities in three of its services (\ie Activity Manager, Location Manager, Mount Service), and two apps from the {\tt Applications} layer (\ie SystemUI, Phone). 

\subsubsection{Mining-Based Studies} Closer to our study are the previous works aimed at analyzing security vulnerabilities by following a mining-based approach. Some of those studies are Android-specific  \cite{Thomas:SPSM2015,Thomas:2015,Jimenez:QRS16} while others are more general in the sense that they aim at characterizing security bugs \cite{Zaman:MSR11,Lal:APSEC12}. Thomas \etal \cite{Thomas:SPSM2015} mined the OS updates installed on 20k+ Android devices to measure the delivery time of security updates for 11 vulnerabilities, and to establish a scoring model of insecure devices; the results suggest that, on average, 87.7\% of the devices are exposed to at least one of the analyzed vulnerabilities. Thomas \cite{Thomas:2015} investigated the CVE-2012-6636\cite{CVE-2012-6636} vulnerability on the JavaScript-to-Java interface of the WebView API; 102k+ APKs were statically analyzed to measure the number of apps in which the vulnerability could be exploited. In addition, the lifetime of the vulnerability was analyzed using an approach similar to \cite{Thomas:SPSM2015}. 

Finally, Jimenez \etal \cite{Jimenez:QRS16} analyzed 32 vulnerabilities from the CVE database\cite{CVE} to identify the issues, involved components, code complexity of the patches, and complexity of the code methods/functions involved in the vulnerability.  

The study presented in this paper is complementary to previous studies in the sense that a larger set of vulnerabilities (mined from CVE) is analyzed (660) and different perspectives are included in the study such as the survivability time of the vulnerabilities, subsystems and components of the Android OS involved in the vulnerabilities, an extensive taxonomy of security issues based on the Common Weakness Enumeration (CWE) hierarchy of vulnerabilities\cite{CWE}, and a list of learned lessons oriented to Android OS and apps developers.

\section{Study Design}
\label{sec:design}
\newcommand\rqone {Which types of security vulnerabilities affect Android?}
\newcommand\rqtwo {Which are the Android subsystems more affected by security vulnerabilities?}
\newcommand\rqthree {How long does it take to fix security vulnerabilities in Android?}

The {\em goal} of the study is to investigate Android-related security vulnerabilities reported over the past eight years (\ie the whole history of the Android OS). The {\em purpose} is to (i) define a taxonomy highlighting the types of Android-related vulnerabilities as well as which of the Android OS subsystems are more exposed to security issues, and (ii) investigate the time needed to fix vulnerabilities in Android. 
The {\em context} consists of 660 vulnerabilities mined from CVE Details\cite{CVE, CVE-Android}, a vulnerability datasource processing XML feeds provided by the National Vulnerability Database\cite{NVD-RSS}. All the data used in the study are available in our online appendix \cite{replication}.

\begin{figure}[t]
\begin{center}
\includegraphics[width=\linewidth]{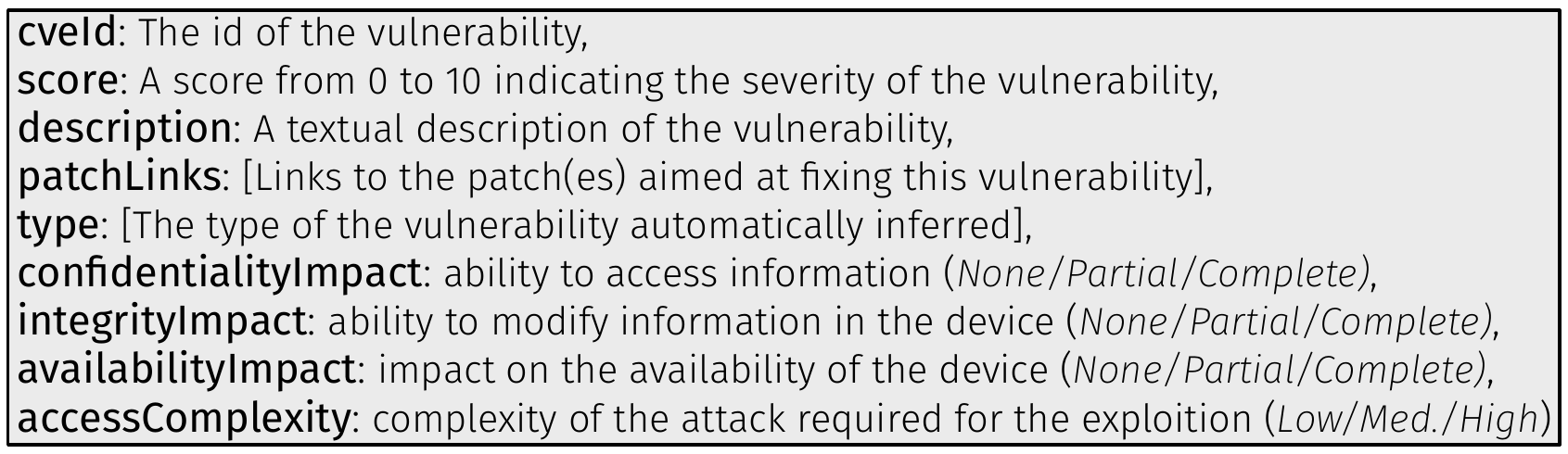}\vspace{-0.3cm}
\caption{Information stored for the vulnerabilities mined from CVE Details}\vspace{-0.7cm}
\label{fig:json}
\end{center}
\end{figure}

\noindent The study addresses the following research questions:

{\bf RQ$_{1}$:} {\em \rqone} This research question aims at identifying the types (\eg inadequate encryption strength) of Android-related vulnerabilities reported over the past eight years. Note that with ``Android-related'' we refer to both vulnerabilities directly affecting code components belonging to the Android OS, \ie the components of the Android software stack developed by Google, as well as those related to third-party components (\eg hardware drivers, apps shipped with the devices) threatening the security of Android devices. Also, we investigate (i) the impact on \emph{confidentiality}, \emph{integrity}, and \emph{availability} of the vulnerabilities (see \figref{fig:json} for a definition of these three properties), and (ii) the complexity of the attack required to exploit the vulnerabilities (\emph{accessComplexity} in \figref{fig:json}). %Knowing the types of security vulnerabilities affecting the Android devices and their characteristics can help to guide (i) apps developers, in focusing their verification \& validation activities toward the identification of the most frequently reported types of vulnerability,  (ii) researchers, in investing in vulnerabilities detection tools targeting the most diffused types of Android-related vulnerabilities (thus being particularly valuable to increase the security of Android devices), and (iii) language/API developers, to design/improve mechanisms for secure coding of Android apps and the underlying platform. 

{\bf RQ$_{2}$:} {\em \rqtwo} The second research question sheds the light on the Android subsystems more frequently affected by security vulnerabilities. Note that in this case our focus will be on the architecture of the Android OS, while less emphasis will be given to vulnerabilities affecting third-party components. Indeed, our goal is to point out to developers (both apps developers as well as contributors of the Android OS) which are the more risky services, APIs, apps, \etc, in the OS. This information can be used to better focus verification \& validation activities as well as to develop better Android-specific tools for vulnerability detection and secure coding.

{\bf RQ$_{3}$:} {\em \rqthree} This research question studies the survivability of the security vulnerabilities subject of our study. In particular, we assess the number of days between the vulnerability introduction and its fixing. RQ$_{3}$'s findings could help in assessing the usefulness of effective vulnerability detection tools able to immediately catch an introduced vulnerability (\ie a long survivability of the vulnerabilities would indicate the urge for such tools), and to identify the prevalence of vulnerabilities across different versions of the OS. 

%\item {\bf RQ$_{4}$:} {\em \rqfour} This last question aims at ``profiling'' the developers responsible for introducing and fixing security vulnerabilities. In particular, we verify if the developer's experience influences the likelihood for a developer to be responsible for the introduction and/or fixing of vulnerabilities. This information, combined with the RQ$_{2}$'s findings, could provide indications on the experience of the developers that should be allocated for implementation tasks impacting vulnerability-prone subsystems.

\subsection{Data Extraction and Analysis}
\label{sub:context}
The context of the study consists of 660 vulnerabilities mined from the official Android Security Bulletins and the CVE Details website\cite{CVE}. The information was collected on November 24, 2016. First, we built a web-based scraper that went over all the 16 bulletins published by Google from August 2015 until November 2016, looking for CVE ids (using regular expressions). In total, we found 564 CVE ids in the Android Bulletins; 62 of them are reported as reserved, meaning that the details of the vulnerability are not publicly available. 

A second web scraper was then used to automatically extract the details of each of the vulnerabilities listed in CVE details under the category ``Android''~\cite{CVE-Android}; we collected 629 vulnerabilities from CVE details under the ``Android'' category.  

Some of the non-reserved vulnerabilities (listed in the bulletins) do not appear tagged as Android-related in the CVE details website because they affect the Linux Kernel or third party components (\eg drivers). We found 31 vulnerabilities in the bulletins not tagged as Android-related in CVE details.  Therefore, using the CVE ids from the bulletins we directly scraped the information from CVE details, without relying on the Android filter. At the end, we obtained information for a total of 660 vulnerabilities (629 tagged as Android-related + 31 not tagged as Android but listed in the bulletins). Note that 159 of the collected 660 vulnerabilities are not listed in the bulletins; those CVEs are mostly from drivers, apps, and OS modifications of device manufacturer/vendors, and vulnerabilities found before the first Google bulletin was published (August 13, 2015).

For each of the selected vulnerabilities we stored a {\tt JSON} file reporting the information detailed in \figref{fig:json}. This data was complemented/fixed via manual analysis (as detailed below), and then used to answer RQ$_{1}$ and RQ$_{2}$.  In particular, once extracted the information in \figref{fig:json} for each vulnerability, two authors manually analyzed each vulnerability to: 

\emph{1. Check and complement the vulnerability type automatically inferred by CVE Details, and obtain its hierarchy}. CVE Details exploits a keywords-based mechanism to automatically infer the type of each vulnerability according to the Common Weakness Enumeration (CWE) dictionary \cite{CWE}. Such an automatic process can introduce imprecisions in the data. For this reason, two authors analyzed all the information available about each vulnerability (\ie its page on the National Vulnerability Database, fixing patches when publicly available, official vulnerability bulletins, the Android issue tracker, \etc) to verify the type of the vulnerability, identify the CWE hierarchy, and consequently change/complement the classification (still according to the CWE dictionary). Note that a vulnerability can belong to multiple types having hierarchical relationships between them. For example, a vulnerability can be classified as (from the least to the most specialized category): 
\begin{myquote}
\emph{Improper Restriction of Operations within the Bounds of a Memory Buffer $\rightarrow$ Out-of-bounds Read $\rightarrow$ Buffer Over-read} 
\end{myquote}
Overall, the manual analysis led to the change or complementing (\ie multiple types are assigned to the vulnerability, including the one automatically inferred by CVE Details) of the type provided by CVE Details for 68\% of the analyzed vulnerabilities. 

%\emph{2. Check and complement the ``result of the vulnerability exploitation'' reported in CVE Details}. Also for this information the two authors performed a manual checking by relying on the same set of documents analyzed in the previous step (\ie the Android issue tracker, \etc).  Again, the reported information was corrected and/or complemented (a vulnerability exploitation could have multiple effects, \eg \emph{gain privileges} and \emph{obtain information}). Through this process the authors fixed/complemented this specific information in 10\% of the analyzed vulnerabilities. 

\emph{2. Identify the subsystems affected by the vulnerability}. The authors analyzed the information in the National Vulnerability Database (including, when available, the patches fixing the vulnerability) as well as online documentation (\eg the Android issue tracker) to identify the code components affected by the vulnerability. Firstly, a high-level classification was performed (\ie the vulnerability affects the Android OS components developed by Google or third-party components). 

Then, for the vulnerabilities affecting the Android OS, a more fine-grained category was defined in order to identify the affected architectural layer (\eg \emph{Android runtime}) and, more specifically, the affected subsystem (\eg \emph{Dalvik VM}). %The considered architectural layers and subsystems are presented in our results discussion. 

The above described manual analysis was performed in three rounds. First, two authors ($A_1$ and $A_2$) manually analyzed half of the 660 vulnerabilities each. Then, $A_1$ checked the vulnerability types and the impacted architectural layers/subsystems assigned by $A_2$ and \emph{vice versa}. Finally, the authors discussed the 47 (7\%) cases of disagreement, reaching an agreement on the correct classification needed. One vulnerability ({\tt CVE-2016-3877}) has been excluded from the study at this stage, since no information was available about it. Also, in cases in which the two evaluators were undecided about the specific type of vulnerability and/or about the subsystem affecting the vulnerability, an ``unclear'' tag was assigned. 

We answer RQ$_{1}$ by presenting a taxonomy of the types of vulnerabilities identified in the manual analysis as well as descriptive statistics about their characteristics (\eg impact on \emph{confidentiality}). The characteristics have not been manually validated since they are mined by CVE Details directly from the National Vulnerability Database\cite{NVD-RSS}. Thus, we assume them as correct. Concerning RQ$_{2}$, we report a heat map showing the distribution of vulnerabilities across the Android subsystems. We complement our discussion with qualitative examples.

To answer RQ$_{3}$ we need information not available in the CVE Details datasource. In particular, we need to identify the commits in which each vulnerability has been introduced and fixed. As for the commit fixing each vulnerability, we mined it from the Android Security Bulletins\cite{AndroidSecurityBulletins}, issued each month and reporting about the recently identified/fixed vulnerabilities. %We mined the security bulletins from August 2015 (\ie the first available) to November 2016. 
The vulnerability-fixing commit is not available for all the 660 Android-related vulnerabilities we collected from the CVE Details datasource, because (i) some vulnerabilities were reported before the first available bulletin, and (ii) the fixing commit (see \eg \url{http://tinyurl.com/hrod7q9}) is only available for the subset of vulnerabilities fixed in the Android open source project (\eg it is not available for vulnerabilities related to third-party components such as drivers). Note also that, although the CVE reports include in some cases the bug id, the ids are for the internal bug trackers of Google and hardware manufactures, which are not publicly available. For these reasons, the analysis for RQ$_{3}$ is limited to a set of 201 vulnerabilities for which we identified the fixing commit. Once identified the fixing commit, we used the SZZ algorithm \cite{SZZ} to identify the commit introducing the vulnerability. The algorithm relies on the annotation/blame feature of versioning systems. Given a vulnerability-fixing commit VF$_k$ (where $k$ identifies the vulnerability), the approach works as follows:

1. For each file $f_i$, involved in VF$_k$ and fixed in its revision {\em rel-fix}$_{i,k}$, we extract the file revision just {\em before} the vulnerability fixing ({\em rel-fix}$_{k}-1$). %\MARIO{Gabriele, do we need the index i in {\em rel-fix}$_{i,k}$? it suggest that there is a vulnerability fix for each file $f_i$. Is that right? } YES

2. Starting from the revision {\em rel-fix}$_{k}-1$, for each source code line in $f_{i}$ changed to fix the vulnerability $k$, the {\em blame} feature of Git is used to identify the file revision where the last change to that line occurred. 

In doing that, blank lines and lines that only contain comments are identified using an island grammar parser \cite{Moonen:wcre2013}. This produces, for each file $f_i$, a set of $n_{i,k}$ fix-inducing revisions {\em rel-vulnerability}$_{i,j,k}$, $j=1 \dots n_{i,k}$.  

Since more than one commit can be indicated by the SZZ algorithm as responsible for inducing the vulnerability-fix, there are time vulnerability ranges defined by lower (minimum survivability) and upper bounds (maximum survivability). Therefore, we answer RQ$_{3}$ by following a meta analysis-based procedure\cite{Hedges:1985,Cumming:2011}: The minimum and the maximum survivability of the vulnerabilities (\ie number of days between the vulnerability introduction and fixing) are plotted using forest plots with confidence intervals, and a central tendency measure of the survivability is computed by using the random effects model \cite{Cumming:2011} (based on the recommendations by Cumming\cite{Cumming:2011}). The minimum survivability is the one observed when considering the most recent commit identified by the SZZ algorithm as the one that induced the vulnerability-fix. \emph{Vice versa}, the maximum survivability is observed when considering the least recent commit identified by the SZZ algorithm as the one that induced the vulnerability-fix. The forest plots are depicted by considering a 95\% confidence interval. 

We also verify whether vulnerabilities having different severity levels have different survivability. For this analysis, we use the severity classification available in the Android bulletins (\emph{low}, \emph{moderate}, \emph{high}, and \emph{critical}). In particular, we compare the distributions of the survivability of the different categories of vulnerabilities (\eg \emph{low vs. moderate}) via (i) forest plots, and (ii) statistical tests. For the latter we exploit the Mann-Whitney test \cite{Conover:1998} with results intended as statistically significant at $\alpha = 0.05$. To control the impact of multiple pairwise comparisons (\eg the survivability of the vulnerabilities having \emph{low} severity is compared against the survivability of those having \emph{moderate}, \emph{high}, and \emph{critical} severity), we adjust $p$-values using the Holm's correction \cite{holm}. We also estimate the magnitude of the differences by using the Cliff's Delta ($d$), a non-parametric effect size measure \cite{Cliff:2005} for ordinal data. We follow well-established guidelines to interpret the effect size: negligible for $|d| < 0.10$, small for $0.10 \le |d| < 0.33$, medium for $0.33 \le |d| < 0.474$, and large for $|d| \ge 0.474$ \cite{Cliff:2005}.

\section{Results}
\label{sec:results}
This section discusses the quantitative results achieved in our study according to the three formulated RQs. Also, we complement quantitative data with qualitative examples, by referring to specific vulnerabilities identified with their CVE id (\eg {\tt CVE-2016-2439}). The reader can access the page detailing a vulnerability by visiting ``\url{https://web.nvd.nist.gov/view/vuln/detail?vulnId=}'' followed by the CVE id. Due to the lack of space we only discuss a limited set of examples. However, in our online appendix \cite{replication} we provide the complete list of vulnerabilities considered in our study, including their categorization by subsystem, component, and vulnerability type. Also, we created visualizations aimed at helping the reader when browsing the vulnerabilities list \cite{replication}. 

\begin{sidewaysfigure*}[!]
    \includegraphics[width=\hsize]{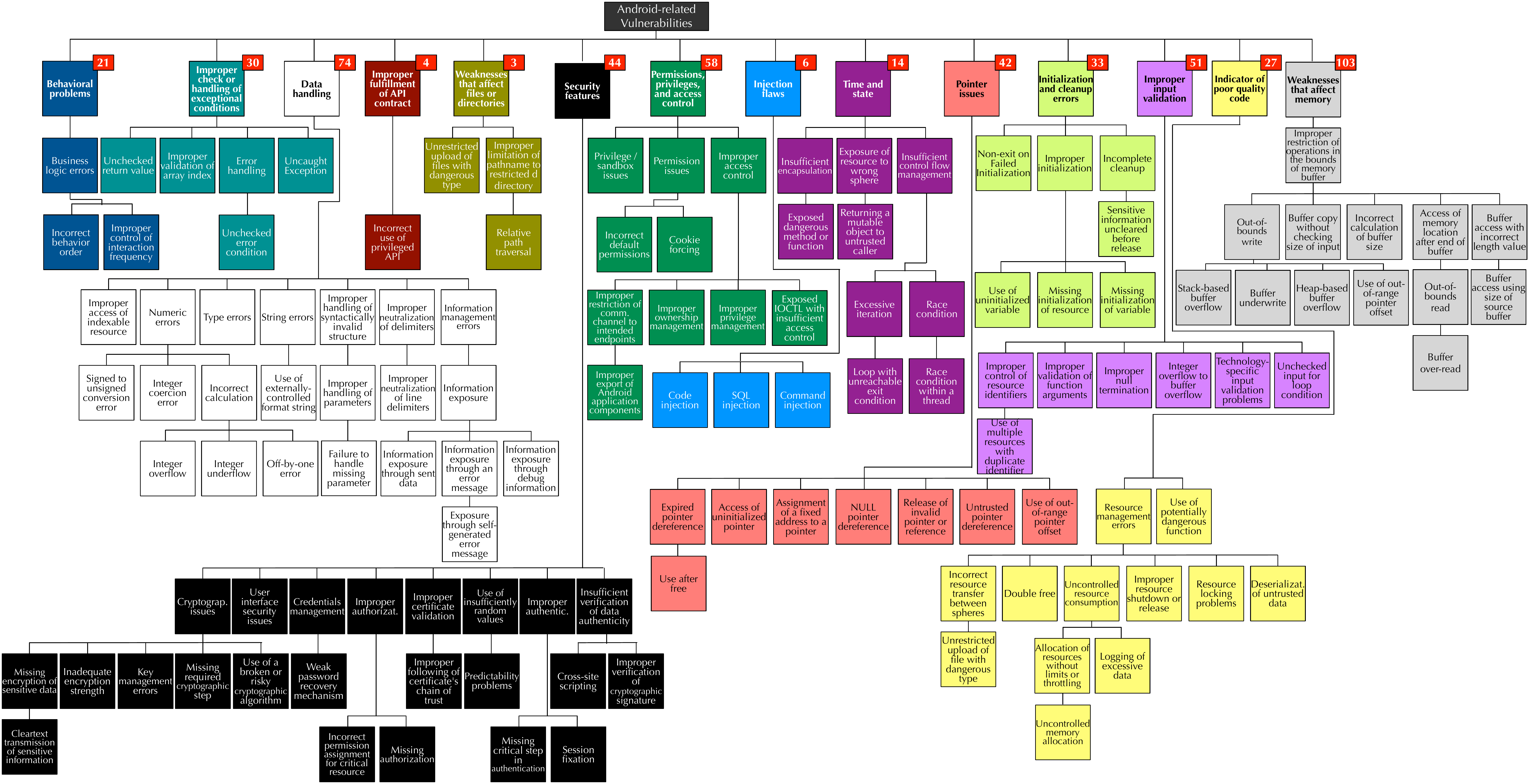}
    \caption{RQ$_{1}$: Types of Android-related vulnerabilities}
    \label{fig:types}
\end{sidewaysfigure*}

\subsection{\rqone}
\figref{fig:types} shows the taxonomy reporting the vulnerability types we found in the 660 manually inspected vulnerabilities. As explained in \secref{sub:context}, the vulnerability types are depicted in a hierarchical manner by following the categorization provided in the CWE dictionary \cite{CWE}. Note that \figref{fig:types} only reports the classification for 510 vulnerabilities. This is due to the fact that we were not able to infer the type of 150 vulnerabilities during our manual analysis.

The vulnerabilities most frequently affecting Android devices are those related to \textbf{weaknesses that affect the memory}, with 103 instances (20\%). These weaknesses include all vulnerabilities related to the \emph{improper restriction of operations in the bounds of memory buffer}, like \emph{out-of-bounds read/write}. One vulnerability falling in this category is {\tt CVE-2016-2439}, described as follows:
\begin{myquote}
\emph{Buffer overflow in btif\_dm.c in Bluetooth [...] allows attackers to execute arbitrary code via a long PIN value}
\end{myquote}
The vulnerability was fixed in commit {\tt 9b534de}, modifying the conditional statement checking a PIN-related error from {\tt if (pin\_code == NULL)} to {\tt if (pin\_code == NULL || pin\_len $>$ PIN\_CODE\_LEN)}. The rationale behind the change is also documented by the developer in the commit message: \emph{If a malicious client set a pin that was too long it would overflow the pin code memory}.

Very popular are also vulnerabilities related to \textbf{data handling}, typically found in functionalities that process data \cite{CWE} (74 instances---15\%). These include, for example, \emph{type errors} like the one related to {\tt CVE-2016-3918}:
\begin{myquote}
\emph{AttachmentProvider.java in AOSP Mail in Android [...] does not ensure that certain values are integers, which allows attackers to read arbitrary attachments [...]}
\end{myquote}
The vulnerability was fixed in commit {\tt 6b2b0bd} that, as reported in the commit message: \emph{Limits account id and id to longs [...] Both id and account id are now parsed into longs (and if either fails, an error will be logged and null will be returned)}. Note that the  \emph{data handling} category includes several different sub-categories that we do not detail due to lack of space (see \figref{fig:types}). %The complete categorization is available in our online appendix \cite{replication}.

Vulnerabilities related to \textbf{permissions, privileges, and access control} are represented in our taxonomy with 58 instances (11\%). They include, for example, weaknesses due to \emph{improper access control} and to \emph{permission issues}, like the \emph{cookie forcing} vulnerability discussed in {\tt CVE-2008-7298}:
\begin{myquote}
\emph{The Android browser cannot properly restrict modifications to cookies established in HTTPS sessions, which allows man-in-the-middle attackers to overwrite or delete arbitrary cookies via a Set-Cookie header in an HTTP response [...]}
\end{myquote}

\textbf{Improper input validation} (51 instances---10\%) includes vulnerabilities caused by a missing or improper validation of inputs that can affect the control/data flow of the program \cite{CWE}. Vulnerabilities in this category include (but are not limited to---see \figref{fig:types}) \emph{unchecked input for loop condition} and \emph{improper validation of function arguments}. 

The latter are the most popular in this category and, while their fixing is generally simple (\eg the addition of a missing/improper argument validation), they can result in severe attacks like the one possible by exploiting {\tt CVE-2016-3910} (9.3 out of 10 in severity score).

\textbf{Security features} are involved in 44 vulnerabilities (9\%) related to \emph{cryptographic issues}, \emph{user interface security issues}, \emph{credentials management} problems, \etc (see \figref{fig:types}). For example, {\tt CVE-2011-2344} reports a vulnerability due to \emph{inadequate encryption strength} possibly causing severe attacks allowing the stealing of private pictures:
\begin{myquote}
\emph{Android Picasa in Android [...] uses a cleartext HTTP session when transmitting the authToken obtained from ClientLogin, which allows remote attackers to gain privileges and access private pictures and web albums by sniffing the token from connections with picasaweb.google.com}
\end{myquote}

\textbf{Initialization and cleanup errors} and \textbf{improper check or handling of exceptional conditions} are the cause for 33 and 30, respectively, of the categorized vulnerabilities ($\sim$6\% each). These categories include, among others, the \emph{missing initialization of a variable} and \emph{uncaught exceptions}.

Finally, other less diffused vulnerabilities are those falling in the categories: \textbf{Indicator of poor quality code} (27 instances), \textbf{behavioral problems} (21), \textbf{time and state} (14), \textbf{injection flaws} (6), \textbf{improper fulfilment of API contract} (4), and \textbf{weaknesses that affect files or directories} (3). A description of these categories can be found in the CWE dictionary \cite{CWE}, while Android-related examples from our dataset are available in our online appendix \cite{replication}.

\subsubsection{Characteristics of the Android-related vulnerabilities} 
We discuss the characteristics of the vulnerabilities in terms of their access complexity and availability, integrity, and confidentiality impact (see \figref{fig:json} for a definition of these characteristics).

\begin{figure}[t]
\begin{center}
\includegraphics[width=0.7\linewidth]{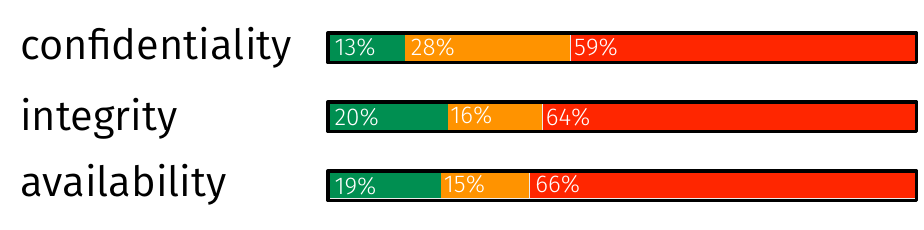}\vspace{-0.3cm}
\caption{RQ$_{1}$: Impact of the vulnerability exploitation}\vspace{-0.7cm}
\label{fig:impacts}
\end{center}
\end{figure}

\noindent \textbf{Access complexity.} Very few vulnerabilities (21) require a high access complexity, meaning that (i) the vulnerability is difficult to exploit, and (ii) specific conditions must verify to allow the exploitation. Most of the vulnerabilities have either a medium (399) or low (238) access complexity. Note that having a low access complexity (\ie very little knowledge needed to exploit the vulnerability) does not imply a lower severity for the effects for the exploitation. Indeed, 130 of these vulnerabilities in our dataset (\ie 55\% of all those having a low access complexity) cause a complete \emph{confidentiality} (\ie total information disclosure), \emph{integrity} (\ie total compromise of system integrity), and \emph{availability} (\ie total unavailability of the targeted device resources, like CPU) impact. 

\noindent \textbf{Impact of the exploitation.} \figref{fig:impacts} reports the impact on confidentiality, integrity, and availability of the studied vulnerabilities. Green indicates \emph{no impact}, orange a \emph{partial impact}, and red a \emph{complete impact}. 

Most of the Android-related vulnerabilities can seriously compromise the confidentiality and integrity of the information stored in the device and can cause a complete exhaustion of the device's resources. This highlights the urge for techniques and tools supporting the detection of Android-related vulnerabilities at different stages: development,  submission to the market, and execution in the device. 

\begin{figure}[t]
\begin{center}
\includegraphics[width=0.8\linewidth]{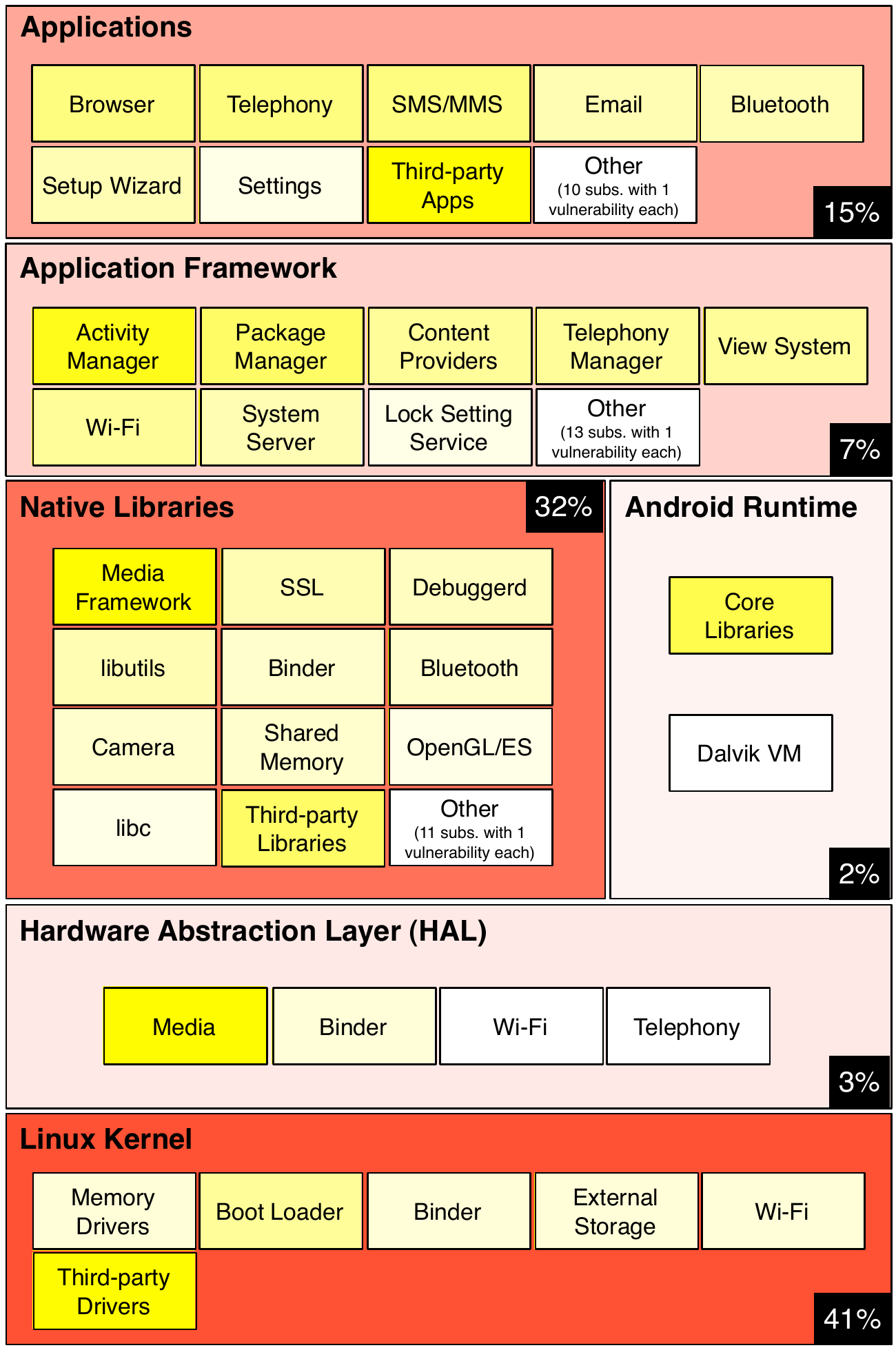}\vspace{-0.3cm}
\caption{RQ$_{2}$: Heat map of vulnerabilities in the Android layers/subsystems}\vspace{-0.7cm}
\label{fig:heatmap}
\end{center}
\end{figure}

\subsection{\rqtwo}
\figref{fig:heatmap} depicts (using a heat-map style), the manually identified layers and subsystems of the Android OS impacted by 634 vulnerabilities. For building the heatmap, from the 660 vulnerabilities dataset, we excluded 26 in which we were not able to manually identify the layer. For the heat-map we used two color schemes: white-to-red for the layers, with white representing the lowest value and red the highest one; and white-to-yellow for the subsystems (\ie internal boxes), with full yellow meaning that a subsystem is responsible for 100\% of the vulnerabilities in the corresponding layer. Note that the subsystems' colors are normalized on the basis of the total vulnerabilities affecting a layer. \figref{fig:heatmap} also reports the percentage of vulnerabilities affecting each layer. 

{\tt Linux Kernel} is the most frequently affected layer, with 261 of the 634 vulnerabilities (41\%). It is worth noting that the Android Open Source Project includes modifications to the original kernel to enable mobile features. However, most of the vulnerabilities in this layer affect third-party drivers developed by hardware manufactures (OEMs): 237 vulnerabilities are in third-party drivers, while only 14 are from Google changes/contributions to the kernel and are related to Android-specific components such as {\tt Binder}, {\tt ashmem/Shared memory}, and {\tt aboot/Boot loader}.  For 10 of the 261 vulnerabilities we were not able to identify whether they affect kernel components contributed by Google or by OEMs. For third-party drivers, {\tt Video}, {\tt WiFi}, and {\tt Camera} are the top three hardware components/features involved in the vulnerabilities. In the case of the Android-specific components, most of the reported vulnerabilities (8 out of 14) are located in the {\tt Bootloader}, and correspond to overflow/over-read issues, improper check or handling of exceptional conditions\cite{CWE-703}, improper access control\cite{CWE-284}, and weak password recovery mechanism \cite{CWE-640}.

The {\tt Native libraries} layer is the second one exhibiting the largest number of vulnerabilities (201 out of 634 = 31.7\%). This is mostly due to the {\tt Media Framework} subsystem that has suffered of 143 vulnerabilities (129 from Google contributions, 14 from third party contributions), including the set of issues known as ``Stagefright" \cite{Stagefright,Nickinson:2015,Burgess:2016} that are sourced in the {\tt Stagefright} library (\emph{a.k.a.,} libstagefright). In the case of third-party files, the vulnerabilities are in libraries supporting the {\tt Media framework}, {\tt WiFi}, {\tt Bluetooh}, and {\tt DHCP} services,  and the {\tt Skia} library. Vulnerabilities in the {\tt Media Framework} are mostly related to issues with pointers\cite{CWE-465}, arrays access/writing, and memory management that lead to any type of overflow/underflow when accessing, writing, creating, and copying buffers \cite{CWE-120,CWE-121,CWE-122}, and when performing integer operations \cite{CWE-190}. For instance, the vulnerability {\tt CVE-2015-3834} reported as fixed in the August 2015 bulletin has the following CVE description:

\begin{myquote}
\emph{Multiple integer overflows in the BnHDCP::onTransact function in libstagefright allow attackers to execute arbitrary code via a crafted application that uses HDCP encryption, leading to a heap-based buffer overflow [...]}
\end{myquote}
This vulnerability was fixed with the commit {\tt c82e31a} that modifies the {\tt IHDCP.cpp} file. The lack of buffer size validation when computing a buffer size, was leading to heap-based buffer overflows\cite{CWE-122} when creating an input buffer with the calculated size.
Another example of security issue in the {\tt Media Framework} related to improper validation/restriction of operations within the bounds of a memory buffer is {\tt CVE-2016-0815}:
\begin{myquote}
\emph{The MPEG4Source::fragmentedRead function in MPEG4Extractor.cpp in libstagefright in mediaserver [...] allows remote attackers to execute arbitrary code or cause a denial of service (memory corruption) via a crafted media file [...]}
\end{myquote}
The issue can be summarized as an out-of-bounds write\cite{CWE-787} generated when an array offset goes beyond the buffer size. The vulnerability was fixed in commit {\tt 5403587}.

The {\tt Applications} layer is the top three in the list with 88 vulnerabilities located in 18 applications developed by Google, and 10 third-party apps.  

Concerning the third-party applications, the {\tt Adobe Flash Player} is the most vulnerable app with 29 vulnerabilities; the next more vulnerable app is {\tt Firefox} (4 vulnerabilities); {\tt Nvidia Profiler}, {\tt Widevine QSEE trustzone} and {\tt Samsung OMACP} are  the top three with 3 vulnerabilities each. From the 18  apps developed by Google, {\tt Browser}, {\tt Telephony}, and {\tt SMS/MMS} have been the most vulnerable with 5 vulnerabilities each. 

Vulnerabilities in the {\tt Applications} layer are diverse in terms of types. For example, the Android {\tt Browser} had a vulnerability ({\tt CVE-2011-0680}) impacting 6 different versions of the OS (before 2.3.4) which:

\begin{myquote}
	\emph{[...] allows remote attackers to obtain SD card contents via crafted content:// URIs, related to (1) BrowserActivity.java and (2) BrowserSettings.java [...]}
\end{myquote}
The {\tt CVE-2011-0680} vulnerability is an example of information exposure through sent data\cite{CWE-201} because of the lack of URIs validation in the {\tt Browser} app. 

The next Android OS layer more affected by vulnerabilities is the {\tt Android Framework} with 46 vulnerabilities (7.26\%) mostly located in the {\tt Activity Manager} and {\tt Package Manager}. While the former is in charge of tasks such as intents resolution and app/activity launching, the latter manages information and handle tasks related with the Android packages (\ie apps) installed in the device. Conversely to the {\tt Libraries} layer that exhibits a non-diverse set of vulnerabilities (in terms of the type),  the {\tt Android Framework} has been affected by a diverse set of vulnerabilities including code injection\cite{CWE-94}, overflows \cite{CWE-190}, permission issues\cite{CWE-275}, business logic errors\cite{CWE-840}, missing authorizations\cite{CWE-862},  and use of a risky cryptographic algorithm\cite{CWE-327}, among others. An example of vulnerability in the {\tt Android Framework} from the category ``Business logic errors" is {\tt CVE-2016-2500}:

\begin{myquote}
\emph{Activity Manager in Android [...] does not properly terminate process groups, which allows attackers to obtain sensitive information via a crafted application [...]}
\end{myquote}

This vulnerability was a consequence of an invocation to the {\tt killProcessGroup} method in the {\tt ActivityManagerService.java} file using wrong parameters.  Another example of vulnerability introduced by business logic errors in the  {\tt Android Framework} layer is {\tt CVE-2016-3923}, in particular in the {\tt Accessibility Services} that ``\emph{mishandle motion events, which allows attackers to conduct touchjacking attacks and consequently gain privileges via a crafted application}". 

The {\tt HAL}, and {\tt Android Runtime} layers are the ones less impacted by vulnerabilities with 19 and 14 vulnerabilities, respectively. The interface for media components  is the most impacted component from {\tt HAL}. As for the {\tt Android Runtime}, most of the vulnerabilities affect core libraries such as  {\tt Apache Harmony}, {\tt Bouncy Castle}, and  {\tt Conscrypt}; only two vulnerabilities were reported for the {\tt Dalvik VM}. Finally, 5 out of the 634 vulnerabilities (0.79\%) vulnerabilities (not included in the heatmap) were manually assigned to different layers because the patches modified diferent layers of the Android OS stack; those vulnerabilities are CVE-2016-3760, CVE-2010-4832, CVE-2015-3843, CVE-2016-2496, and CVE-2016-3889.

\begin{figure}[t]
	\begin{center}
		\includegraphics[width=0.91\linewidth]{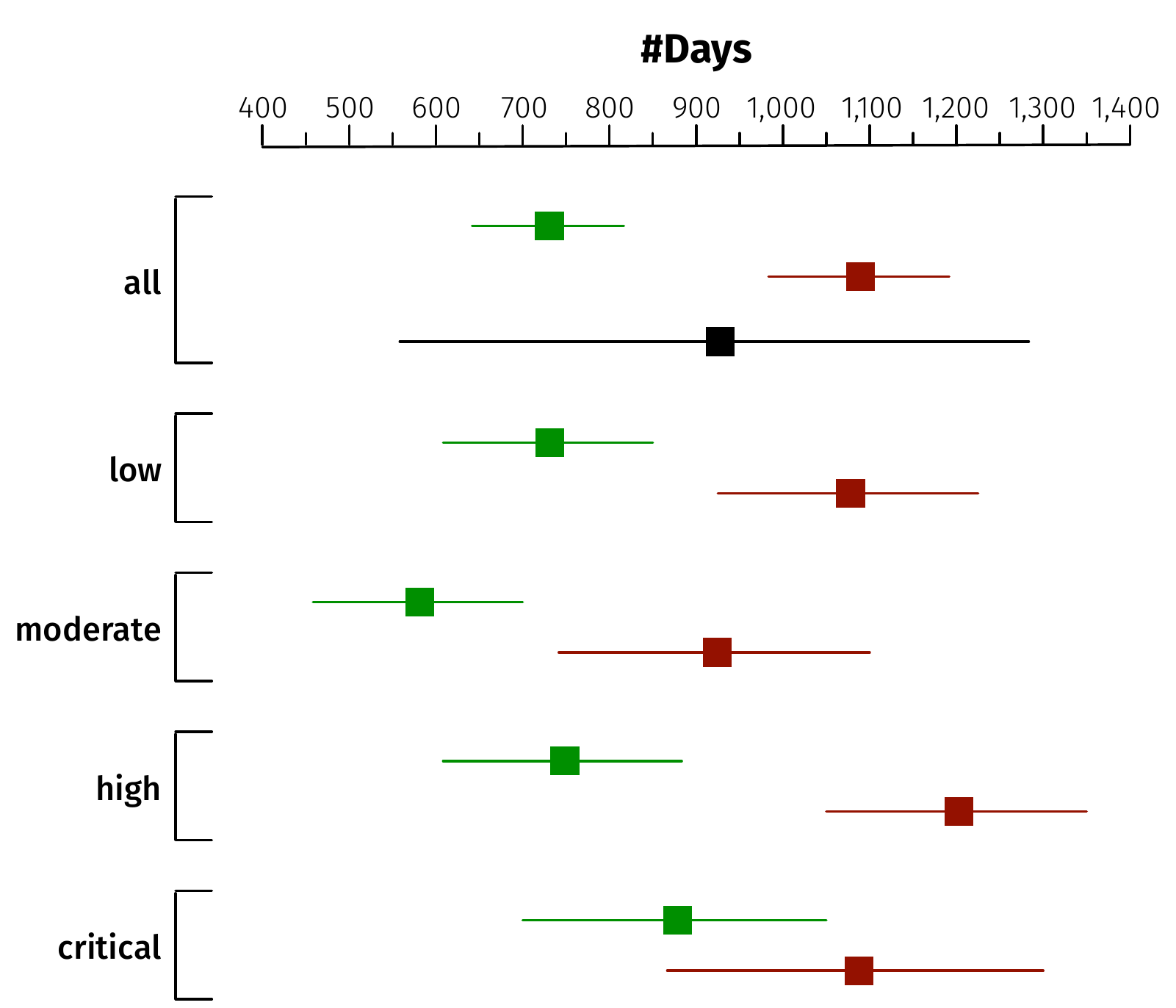}\vspace{-0.3cm}
		\caption{RQ$_{3}$: Survivability in days of Android-related vulnerabilities. Green (red) depicts minimum (maximum) estimates at 95\% confidence interval. Black shows the results of the random effect model.}\vspace{-0.8cm}
		\label{fig:survivability}
	\end{center}
\end{figure}

\subsection{\rqthree}
\figref{fig:survivability} depicts the forest plots reporting the survivability of Android-related vulnerabilities (\ie the number of days between the vulnerability introduction and its fixing). As explained in \secref{sub:context}, we report the minimum (green) and the maximum (red) survivability intervals as computed with the SZZ algorithm. In each forest plot the square represents the average value of the distribution, while the line passing through it depicts the 95\% confidence interval. \figref{fig:survivability} shows the survivability intervals when considering all the analyzed vulnerabilities together (top part of \figref{fig:survivability}) as well as when grouping them by severity (low, moderate, high, and critical). Finally, the black line shown for the overall set of vulnerabilities depicts the results of the random effects model \cite{Ronald:2011}, used in meta-analysis to combine the results of different studies in a single result outcome. In our case, the set of ``different studies'' includes the survivability estimates when considering the minimum (\emph{study I}) and the maximum (\emph{study II}) survivability. 

The first thing that leaps to the eyes from the analysis of \figref{fig:survivability} is the very long survivability of the analyzed Android-related vulnerabilities. Indeed, even when considering the most conservative results (\ie the minimum estimated survivability---green line), the number of days needed to fix an introduced vulnerability is, on average, 724 (it grows to 907 for the random effects model, and to 1,093 for the maximum estimated survivability). It is important to note that this is not the number of days needed to fix a vulnerability after \emph{it has been reported}, but after \emph{it has been introduced}. This means that a vulnerability could remain unnoticed in the system for years before being identified, possibly exploited, and then fixed. While it would have been interesting to also analyze the time actually needed for the vulnerability fixing (\ie the number of days between the vulnerability reporting and fixing), we did not find a way to reliably identifying the reporting date. 

This very long survivability of the Android-related vulnerabilities was surprising for us at a first sight, especially due to the young age of the Android OS. Thus, we manually inspected twenty randomly selected vulnerabilities in order to verify whether strong imprecisions of the SZZ algorithm were there affecting our findings. Note that such a sample is not statistically significant, but just meant to show qualitative examples about the extracted data. 

Overall, we found the estimates provided by the SZZ algorithm to be precise. In particular, in 13 of the inspected cases the SZZ identified a single commit as the vulnerability-fix-inducing one. In all these cases the identified commit was correct. In the remaining seven cases, multiple commits were identified by the SZZ algorithm as the possible responsible for the vulnerability introduction. In all these cases, either the minimum or the maximum vulnerability estimate was correct. In the following, we discuss some examples of manually inspected vulnerabilities. 

The vulnerability {\tt CVE-2015-1538} has been reported in the August 2015 security bulletin and is described as follows: 
\begin{myquote}
\emph{Integer overflow in the SampleTable::setSampleToChunkParams function in libstagefright in Android before 5.1.1 LMY48I allows remote attackers to execute arbitrary code [...]}
\end{myquote}
Such a vulnerability has been fixed in the commit {\tt cf1581c} made on the 8th April 2015, having commit message \emph{Fix several ineffective integer overflow checks} and modifying the file {\tt libstagefright/\-SampleTable.cpp}. By inspecting the diff of such a commit, three lines were changed to fix the integer overflows\cite{cf1581c66c2ad8c5b1aaca2e43e350cf5974f46d}. The SZZ algorithm correctly identifies the commit {\tt edd4a76e} performed on the 28th July 2014 as the vulnerability-inducing commit (thus, the vulnerability survived in the system for 254 days). Indeed, in such a commit the three lines causing the integer overflows and then fixed were introduced all together, as it can be seen from the commit diff\cite{edd4a76eb4747bd19ed122df46fa46b452c12a0d}. Note that this is one of those cases in which the SZZ algorithm identified a single commit as the responsible for inducing the vulnerability-fix. This was the case for 110 out of the 201 vulnerabilities (55\%) considered in RQ$_{3}$.

For the vulnerability {\tt CVE-2015-6608} we identified instead multiple commits as the possible responsible for the vulnerability introduction. This vulnerability is described as follows:
\begin{myquote}
\emph{[...] allows remote attackers to execute arbitrary code or cause a denial of service (memory corruption) [...]}
\end{myquote}
The vulnerability has been fixed in the commit {\tt 8ec845c} (commit note: \emph{stagefright: check IMemory::pointer() before using the allocation}) made on the 15th May 2015 and modifying two lines\cite{8ec845c8fe0f03bc57c901bc484541bdd6a7cf80} in {\tt media/libstagefright/ACodec.cpp}. These two lines were modified for the last time by two different commits, one performed on the 21st February 2012 (\ie {\tt 5778822}) and one performed on the 2nd May 2013 (\ie {\tt 054e734}). Each of these commits introduced one of the two lines then fixed in {\tt 8ec845c} thus, they were both correctly identified as vulnerability-inducing commits. 

In this case, the commit {\tt 054e734} contributes to the ``minimum survivability distribution'' depicted in green in \figref{fig:survivability} (the survivability is 742 days), while {\tt 5778822} contributes to the ``maximum survivability distribution'' depicted in red in \figref{fig:survivability} (survivability=1,179 days). Clearly, in this case the correct survivability estimate is 1,179, since the vulnerability was there (at least in part) since the 21st February 2012. 

%\begin{table}[tb]
%\caption{RQ$_{3}$: Survivability of vulnerabilities having different severity levels: Mann-Whitney test (adj. $p$-value) and Cliff's Delta ($d$).\vspace{-0.2cm}}
%\label{tab:stats-rq3}
%\centering
%\resizebox{0.7\linewidth}{!}{
%\begin{tabular}{lrr}\hline
%\multicolumn{3}{c}{\textbf{Minimum Estimates}}\\\hline
%Test & adj. $p$-value & $d$ \\ \hline
%low $vs$ moderate & 0.35 & -0.16 (Small)\\
%low $vs$ high & 0.84 & 0.02 (Negligible)\\
%low $vs$ critical & 0.44 & 0.15 (Negligible)\\
%moderate $vs$ high & 0.20 & 0.20 (Small)\\
%moderate $vs$ critical & 0.06 & 0.29 (Small)\\
%high $vs$ critical & 0.44 & 0.14 (Negligible)\\\hline
%
%\multicolumn{3}{c}{\textbf{Maximum Estimates}}\\\hline
%Test & adj. $p$-value & $d$ \\ \hline
%low $vs$ moderate & 0.80 & -0.12 (Negligible)\\
%low $vs$ high & 0.77 & 0.12 (Negligible)\\
%low $vs$ critical & 0.89 & -0.01 (Negligible)\\
%moderate $vs$ high & 0.06 & 0.25 (Small)\\
%moderate $vs$ critical & 0.80 & 0.10 (Negligible)\\
%high $vs$ critical & 0.80 & -0.12 (Negligible)\\\hline
%\end{tabular}
%}
%\vspace{-0.5cm}
%\end{table}

When looking for the survivability of vulnerabilities having different severity levels, we were not able to identify any clear trend: It is not possible to assert that vulnerabilities having a higher severity have a higher/lower survivability with respect to those having a lower severity (or \emph{vice versa}). This is visible both from the forest plots (see \figref{fig:survivability}) and confirmed by the statistical analysis, in which we did not observe any significant difference (all the adjusted $p$-values were higher than 0.05).

%Finally, as explained in \secref{sub:context}, we also analyzed the survivability of the vulnerabilities in terms of the number of commits required from their introduction to their fixing. Due to space limitations, the results are not reported here but available in our replication package \cite{replication}. The findings provided by this additional analysis confirmed the very long survivability of the Android-related vulnerabilities, with an average number of commits needed to remove them going from 17,497 (minimum estimate) to 29,250 (maximum estimate). Note that such a high number is the simply result of the long survivability discussed in terms of days multiplied by the high number of commits typical of very active open source projects like Android.

\section{Threats to Validity}
\label{sec:threats}
%This section describes the threats that can affect the validity our study.

Threats to \emph{construct validity} concern the relation between the theory and the observation, and in this work are mainly due to the measurements we performed. This is the most important kind of threat for our study, and is related to:

\emph{RQ$_1$ and RQ$_2$: Subjectivity in the manual classification}. We identified through manual analysis the types of vulnerabilities (RQ$_1$) and the subsystems (RQ$_2$) they affect. To mitigate subjectivity bias in such a process, two authors (A$_1$ and A$_2$) manually analyzed half of the vulnerabilities each. Then, A$_1$ checked the vulnerability types and the impacted subsystems assigned by A$_2$ and \emph{vice versa}. Finally, the authors discussed the cases of disagreement, reaching an agreement on the correct classification needed. Also, when the type of the vulnerability and/or the impacted subsystem was unclear, we preferred to exclude the vulnerability from the study rather than risking to introduce imprecisions.

\emph{RQ$_3$: Approximations due to identifying bug-inducing commits using the SZZ algorithm \cite{Sunghun:tse2011}}. We used heuristics to limit the number of false positives, for example excluding blank and comment lines from the set of bug-inducing changes. Also, we computed both the minimum and the maximum survivability estimates on the basis of the SZZ outcome, showing that in any case the main outcome of our study did not change: Android-related vulnerabilities survive for long time. Moreover, the manual analysis performed on some vulnerabilities confirmed the validity of our experimental design to assess the survivability of vulnerabilities.

\emph{RQ$_3$: Imprecision due to tangled code changes \cite{Herzig:msr2013}}. We cannot exclude that some vulnerability-fixing commits grouped together tangled code changes, of which just a subset was focusing on the vulnerability fix. This would result in imprecisions when running the SZZ algorithm on the fixing commit. Again, by presenting both the minimum and the maximum survivability estimates such a risk is mitigated.

Threats to \emph{internal validity} concern external factors we did not consider that could affect the variables and the relations being investigated. When analysing the survivability of vulnerabilities (RQ$_{3}$) we considered the severity of the vulnerability as a confounding factor to be controlled. 

We are aware that many other factors could influence the survivability, and we plan to analyze them in future work. To reinforce the internal validity, when possible, we integrated the quantitative analysis with a qualitative one.

Threats to \emph{conclusion validity} concern the relation between the treatment and the outcome. Although this is mainly an observational study, wherever possible we used an appropriate support of statistical procedures, integrated with effect size measures that, besides the significance of the differences found, highlight the magnitude of such differences.

Threats to \emph{external validity} concern the generalization of results. In RQ$_1$ and RQ$_2$ we considered 660 vulnerabilities, while the RQ$_3$'s findings  are based on the analysis of 201 vulnerabilities due to the need for identifying the vulnerability-fixing commit (see \secref{sub:context} for details). Clearly, the number of Android-related vulnerabilities that can be studied will increase in the future, and larger replications of our study will be possible.

%\section{Related Work}
%\label{sec:related}
%\input{relatedwork}

\section{Conclusion and Future Work}
\label{sec:conclusion}
We analyzed 660 Android-related vulnerabilities from three different perspectives: (i) the types of the vulnerabilities and their hierarchical relationships, (ii) the layers and components from the Android software stack impacted by the vulnerabilities, and (iii) the survivability of the vulnerabilities (\ie the time required to fix a vulnerability since its introduction). 

The achieved results show that most of the vulnerabilities are related to improper restriction of operations in the bounds of memory buffers, issues processing data (\eg numeric, type, and string errors), improper access control, and improper input validations. This suggests that most of the vulnerabilities can be avoided by relying on secure coding practices especially in the context of data handling and memory access/allocation. Such practices could be enforced, for example, via \emph{just-in-time} quality control techniques statically analyzing the code contributed to the Android OS in each commit activity. Also, mobile OS developers could consider the usage of modern programming languages embedding mechanisms promoting secure coding (\eg Rust \cite{rust}).

Our findings also indicate that third-party hardware drivers are the components mostly affected by security vulnerabilities in the Android OS, thus suggesting the strengthening of verification \& validation activities performed on them. 

Finally, we showed that Android vulnerabilities survive for long time in the code base. This stresses the importance for researchers to invest effort in the development of automatic vulnerability detectors tailored for the mobile world. The taxonomy of vulnerabilities presented in this paper can be used as a reference for the definition of the types of vulnerabilities such detectors should target.  The design and implementation of effective vulnerability detection tools for mobile OS/apps is part of our future research agenda. 

\balance
\bibliographystyle{IEEEtran}
\bibliography{references}

% that's all folks
\end{document}